\documentclass{jpp}

\usepackage{graphicx}
\usepackage{natbib}
\usepackage{amsmath}
\usepackage{color}

\ifCUPmtlplainloaded \else
  \checkfont{eurm10}
  \iffontfound
    \IfFileExists{upmath.sty}
      {\typeout{^^JFound AMS Euler Roman fonts on the system,
                   using the 'upmath' package.^^J}%
       \usepackage{upmath}}
      {\typeout{^^JFound AMS Euler Roman fonts on the system, but you
                   dont seem to have the}%
       \typeout{'upmath' package installed. JPP.cls can take advantage
                 of these fonts, if you use 'upmath' package.^^J}%
      }
  \else
  \fi
\fi

\ifCUPmtlplainloaded \else
  \checkfont{msam10}
  \iffontfound
    \IfFileExists{amssymb.sty}
      {\typeout{^^JFound AMS Symbol fonts on the system, using the
                'amssymb' package.^^J}%
       \usepackage{amssymb}%

      }{}
  \fi
\fi

\ifCUPmtlplainloaded \else
  \IfFileExists{amsbsy.sty}
    {\typeout{^^JFound the 'amsbsy' package on the system, using it.^^J}%
     \usepackage{amsbsy}}
    {\providecommand\boldsymbol[1]{\mbox{\boldmath $##1$}}}
\fi

\newcommand{\fabolt}{f_\alpha(\boldsymbol{r},\boldsymbol{p},t)}
\newcommand{\fakbolt}{f_{\alpha\,\bs{k}}(\boldsymbol{p},t)}
\newcommand{\faksbolt}{f_{\alpha\,\bs{k},\omega}(\boldsymbol{p})}
\newcommand{\faz}{\overline{f}_{\alpha}}
\newcommand{\bs}{\boldsymbol}
\newcommand{\indat}{{\rm init.\,data}}

\title[Collisionless damping of Weibel-generated
  turbulence]{Non-linear collisionless damping of Weibel turbulence in
  relativistic blast waves}

\author[M. Lemoine]%
{M\ls A\ls R\ls T\ls I\ls N\ns L\ls E\ls M\ls O\ls I\ls N\ls E%
  \thanks{Email address for correspondence: lemoine@iap.fr}}

\affiliation{Institut d'Astrophysique de Paris, CNRS - UPMC, 98
  bis boulevard Arago, F-75014 Paris, France
}

\begin{document}

\maketitle

\begin{abstract}
The Weibel/filamentation instability is known to play a key role in
the physics of weakly magnetized collisionless shock waves. From the
point of view of high energy astrophysics, this instability also plays
a crucial role because its development in the shock precursor
populates the downstream with a small-scale magneto-static turbulence
which shapes the acceleration and radiative processes of suprathermal
particles. The present work discusses the physics of the dissipation
of this Weibel-generated turbulence downstream of relativistic
collisionless shock waves. It calculates explicitly the first-order
non-linear terms associated to the diffusive nature of the particle
trajectories. These corrections are found to systematically increase
the damping rate, assuming that the scattering length remains larger
than the coherence length of the magnetic fluctuations. The relevance
of such corrections is discussed in a broader astrophysical
perspective, in particular regarding the physics of the external
relativistic shock wave of a gamma-ray burst.
\end{abstract}


\section{Introduction}\label{sec:intr}
The physics of collisionless shock waves has drawn wide interest, from
pure theoretical plasma physics, starting with the pioneering work of
\citet{1963JNuE....5...43M}, to high energy astrophysics
\citep[e.g.][]{1987PhR...154....1B}, where it plays a key role in
explaining most of the observed non-thermal spectra, and more
recently, to laboratory high energy density physics, where
collisionless shock waves are about to be produced through the
interactions of laser beam-generated
plasmas~\citep[e.g.][]{2012ApJ...749..171D}. At low magnetization --
meaning that the unshocked plasma carries a magnetic field of small
energy density compared to the shock kinetic energy -- the physics of
these collisionless shock waves is driven by the filamentation
instability, also dubbed Weibel instability: this filamentation
instability takes place in the shock precursor, where the incoming
background plasma -- as viewed in the reference frame in which the
shock lies at rest -- mixes with a population of shock-reflected or
supra-thermal particles. This has been demonstrated by ab initio
Particle-in-Cell (PIC) simulations, see
e.g. \citet{2008ApJ...681L..93K} for non-relativistic unmagnetized
shock waves and \citet{2008ApJ...673L..39S} for their relativistic
counterparts, of direct interest to the present work. This
filamentation instability and its various branches have consequently
received a great deal of attention~\citep[see e.g. for relativistic
  shock waves][]{1999ApJ...526..697M, 2004A&A...428..365W,
  2006ApJ...647.1250L, 2007A&A...475....1A, 2007A&A...475...19A,
  2010PhPl...17l0501B, 2010MNRAS.402..321L, 2011ApJ...736..157R,
  2011MNRAS.417.1148L, 2012ApJ...744..182S}.

Further simulations by \citet{2008ApJ...682L...5S} have shown that,
not only can Weibel / filamentation build up the electromagnetic
barrier which gives rise to the shock transition through the
isotropisation of the incoming background plasma, it also builds up
the turbulence which is transmitted downstream of the shock, on plasma
skin depth scales, and which provides the scattering centers for the
Fermi acceleration process. Actually, the excitation of
micro-turbulence -- meaning a turbulence on scales smaller than the
typical gyro-radius of accelerated particles -- is a necessary
condition for a proper relativistic Fermi
process~\citep{2006ApJ...645L.129L,2006ApJ...650.1020N}.

Additionally, \citet{1999ApJ...526..697M} have suggested that the
filamentation mode is able to build up the turbulence in which the
accelerated particles can lose their energy to secondary radiation
through synchrotron (and possibly synchrotron self-Compton)
processes\footnote{Strictly speaking, the relevant radiative processes
  in a microturbulence are jitter and jitter self-Compton, ~\citep[see
    e.g.][]{2011ApJ...737...55M,2013ApJ...774...61K}; however, close
  to the shock front of a relativistic collisionless shock wave, the
  Weibel-generated turbulence is of such strength that the jitter
  radiation boils down to the standard synchrotron spectrum in a
  coherent field of equivalent
  strength~\citep{2009ApJ...707L..92S}. Far from the shock, and in the
  presence of dissipation, jitter effects may in principle become
  significant, depending on how fast the field strength diminishes as
  the effective coherence length grows, see the discussion in
  \citet{2013MNRAS.428..845L}.}. In this unified picture, the
filamentation instability that develops in the shock precursor would
explain a variety of phenomena, from shock formation, to shock
acceleration and even the non-thermal radiation from powerful
astrophysical sources such as gamma-ray bursts. More particularly, the
so-called gamma-ray burst afterglow radiation is attributed to the
acceleration and (synchrotron) radiation of electrons at the external
shock of the gamma-ray burst ultra-relativistic outflow, as it
impinges on the very weakly magnetized circumburst medium. The
phenomenological model of the afterglow provides a satisfactory
description of most observed multi-wavelength afterglow light curves,
see e.g~\citet{2004RvMP...76.1143P}.

A notorious problem of the afterglow model remains to explain the
origin of the magnetic field that permeates the blast, in which the
electrons are assumed to radiate. Indeed, the turbulence which is
generated through the Weibel/filamentation instability in the shock
precursor and transmitted downstream is expected to decay rather
quickly, on multiples of the skin depth scale
\citep{1999ApJ...511..852G}, while the time scales on which the
electrons cool through synchrotron is of the order of
$10^8\,\omega_{\rm p}^{-1}$ for typical external conditions.  This
remark has spurred many theoretical and numerical studies on energy
transfer processes to long wavelengths
\citep[e.g.][]{2005ApJ...618L..75M,2007ApJ...655..375K}, or
alternative instabilities, which might re-amplify the magnetic field
to a fraction of equipartition, from e.g. the interaction of the shock
with an inhomogeneous
medium~\citep{2007ApJ...671.1858S,2014MNRAS.439.3490M}, or from a
Rayleigh-Taylor instability at the contact
discontinuity~\citep{2000astro.ph.12364G,2009ApJ...705L.213L,2010GApFD.104...85L}. How
fast the Weibel-generated turbulence decays thus appears to be a key
ingredient in shaping the light curves of relativistic blast waves.

Recent PIC simulations have addressed this dissipation issue. In PIC
simulations of a relativistic collisionless pair shock up to time
$5300\,\omega_{\rm p}^{-1}$, \citet{2008ApJ...674..378C} have observed
an isotropic, magneto-static turbulence downstream of the shock, which
decays through phase mixing with a damping rate in rough agreement
with the theoretical linear estimate. However, these authors point out
that the linear calculation is ill-suited to describe the damping of
the Weibel-generated turbulence in relativistic blast waves, since the
trajectories of particles deviate from the ballistic regime. This
remark has motivated the present study, which proposes to evaluate the
first non-linear corrections to the damping rate of such
Weibel-generated turbulence, accounting for the deviation of particle
trajectories from straight lines.

The PIC simulations of \citet{2008ApJ...674..378C} have been
essentially confirmed by the more extensive simulations of
\citet{2009ApJ...693L.127K}, although the latter authors observe that
the acceleration of particles to progressively higher energies back
reacts on the structure of the shock, and more importantly, on the
power spectrum of the downstream turbulence, as suggested
independently by \citet{2009ApJ...696.2269M}. Therefore, the former
study concludes that present PIC simulations have not yet converged to
a steady state. Since this longest PIC simulation ($\sim
10^4\,\omega_{\rm p}^{-1}$) represents only a fraction of a percent of
the dynamical timescale of the external shock wave of an actual
gamma-ray burst, while particle acceleration and cooling is believed
to take place on up to this latter timescale, this also means that
theoretical extrapolation is needed to bridge the gap between these
simulations and actual objects. Therefore, the damping rate, which
depends on the power spectrum of the magnetic field, may well differ
from that measured in these PIC simulations. This will be discussed in
some detail further on.

In order to evaluate the non-linear corrections to Landau damping, the
present work calculates the non-linear susceptibility in a
magneto-static turbulence, following the Dupree-Weinstock description
of resonance
broadening~\citep{1966PhFl....9.1773D,1969PhFl...12.1045W,1970PhFl...13.2308W,1975Ap&SS..38..125B}.
This picture has been used in many studies to evaluate the saturation
of instabilities through the back-reaction of particle diffusion in
the grown turbulence, see e.g. \citet{1970PhFl...13.2064D},
\citet{1972PhRvL..28..481B}, \citet{1972PhFl...15..454W},
\citet{1973PhFl...16.2287W} and later works,
e.g. \citet{2011AnGeo..29.1997P} for the particular case of the Weibel
temperature anisotropy. Here, it is used in a different context:
downstream of the shock, the turbulence is magneto-static and
isotropic, therefore the plasma is not subject to any instability,
only to dissipation through phase mixing; the Dupree-Weinstock
approach nevertheless allows to account for the influence of
non-ballistic trajectories on the damping rate. Actually, it will be
shown that a complete calculation of the first order non-linear
corrections is possible, since one can calculate explicitly the
trajectory correlators in a magneto-static small-scale turbulence,
following the method developed in \citet{1977JPlPh..18...49P} and
\citet{2011A&A...532A..68P}.

The results obtained indicate a correction of order unity at the first
non-linear order. However, they also indicate that the correction
systematically increases the damping rate, and that the magnitude of
the correction vs the maximal wavenumber of the turbulence depends on
the power spectrum of the magnetic field. These results are discussed
in a broad context in Sec.~\ref{sec:disc}. Section~\ref{sec:nld}
provides the background for the calculation of the non-linear damping
rate, which is explicitly evaluated in Sec.~\ref{sec:nld-calc}. The
trajectory correlators, which enter the calculation, are discussed in
a separate Appendix~\ref{app:correlators}.

\section{Non-linear damping of small-scale magnetostatic turbulence}\label{sec:nld}
The initial set-up can be described as follows, in the rest frame of
the (downstream) shocked plasma. Time $t=0$ corresponds to the time at
which a given plasma element is advected through the shock towards
downstream; while this plasma element has been crossing the shock
precursor, it has been exposed to micro-instabilities which have built
up a microturbulence to a level characterized by the parameter
$\epsilon_B$:
\begin{equation}
\epsilon_B\,=\,\frac{\left\langle\delta
  B^2\right\rangle}{4\pi\left(\gamma_{\rm rel}-1\right)n m c^2}
\end{equation}
with $m=m_{\rm i}$ for an electron-ion shock, $m=m_e$ for a pair
shock; $n$ represents the particle density in the downstream plasma
rest frame, and $\gamma_{\rm rel}$ represents the Lorentz factor of
the upstream plasma in the downstream rest frame; if $\gamma_{\rm
  sh}\,\equiv\,\left(1-\beta_{\rm sh}^2\right)^{-1/2}$ denotes the
Lorentz factor of the shock front (and $\beta_{\rm sh}$ its velocity
in units of $c$) relatively to the upstream plasma, $\gamma_{\rm
  rel}\,\simeq\,\gamma_{\rm sh}/\sqrt{2}$ for a strong relativistic
weakly magnetized shock~\citep{1976PhFl...19.1130B}. PIC simulations
yield a value $\epsilon_B\,\sim\,10^{-3}-10^{-2}$ immediately
downstream of the shock~\citep{2009ApJ...693L.127K}. The following
calculations describe the microturbulence as aperiodic,
i.e. $\Re\omega\,=\,0$, homogeneous and isotropic, as indicated by PIC
simulations, see in particular~\citet{2008ApJ...674..378C}.

In this respect, the present set-up differs from that of
\citet{2008JETP..107.1049M} and \citet{2010PhRvL.104u5002K} which
derive stationary non-linear and coherent magneto-static solutions to
the Vlasov-Maxwell system in terms of inhomogeneous and anisotropic
particle distribution functions. Such structures indeed emerge in the
shock precursor in the non-linear phase of the instability, as a
balance between the anisotropy/inhomogeneity of the particle
distribution functions and the magnetic forces. In the present case,
the downstream particle distribution function is assumed homogeneous
and isotropic, therefore the plasma is prone to collisionless
damping. The homogeneity and isotropicity of the distribution function
in the downstream plasma is a direct consequence of the shock
transition, as clearly revealed by PIC simulations.

The present microturbulence also differs from the spontaneous
turbulence associated to the thermal fluctuations of the plasma, as
studied recently by ~\citet{2013PhPl...20e2113F},
\citet{2013PhPl...20h2117F}, \citet{2013PhPl...20j4502F},
\citet{2013PhPl...20k2104R} or \citet{2014PhPl...21c2109Y}, since the
present turbulence has been sourced in the shock precursor by the
anisotropies of particle distribution functions.

Finally, the present work neglects any background magnetic field; in
the case of the external shock wave of a gamma-ray burst, this is a
very good approximation, since the magnetization parameter
$\sigma\,\equiv\,B_{\rm ISM\vert d}^2/\left[4\pi \left(\gamma_{\rm
    rel}-1\right)nmc^2\right]$ expressed in terms of the downstream
frame background field $B_{\rm ISM\vert d}$ is very small compared to
$\epsilon_B$: $\sigma\,\sim\,10^{-9}$ for typical interstellar
conditions. Furthermore, the development of the relativistic Fermi
acceleration process requires
$\sigma\,\ll\,\epsilon_B^2$~\citep{2009MNRAS.393..587P,2010MNRAS.402..321L},
i.e. a weakly magnetized shock wave in which the effects of the
background magnetic field can be neglected.

\subsection{Damping of magneto-static turbulence}
Following \citet{2008ApJ...674..378C}, one can use Poynting's theorem to
derive the damping rate as a function of the transverse
susceptibility. For random electric $\bs{\delta E}$ and magnetic
fields $\bs{\delta B}$ and random current density fluctuations
$\bs{\delta j}$ with zero spatial average, Maxwell equations imply
\begin{equation}
\frac{1}{8\pi c}\frac{\partial \delta B^2}{\partial t} +
\frac{1}{8\pi c}\frac{\partial \delta E^2}{\partial t} +
\frac{1}{c}\bs{\delta j}\cdot\bs{\delta E} +
\bs{\nabla}\cdot\left(\bs{\delta E}\times\bs{\delta B}\right)\,=\,0\ .
\end{equation}
Then, taking the average over space, assuming homogeneous turbulence
of strongly magnetic nature, which implies
$\bs{\nabla}\cdot\left\langle\bs{\delta E}\times\bs{\delta
  B}\right\rangle\,=\,0$ and $\delta E^2\,\ll\,\delta B^2$, one
arrives at
\begin{equation}
\frac{1}{8\pi}\frac{{\rm d}\left\langle \delta B^2\right\rangle}{{\rm
    d}t}\,=\,-\left\langle\bs{\delta j}\cdot\bs{\delta E}\right\rangle\ ,
\end{equation}
or
\begin{eqnarray}
\frac{1}{8\pi}\frac{{\rm d}\left\langle\delta B^2\right\rangle}{{\rm
    d}t}&\,=\,&-\frac{1}{2}\int \frac{{\rm d}^3k}{(2\pi)^3}\frac{{\rm
    d}\omega}{2\pi}\, \frac{{\rm d}^3k'}{(2\pi)^3}\frac{{\rm
    d}\omega'}{2\pi}\left\langle\bs{\delta j_{k\omega}}\cdot\bs{\delta
  E_{k'\omega'}^\star}+\bs{\delta j_{k\omega}^\star}\cdot\bs{\delta
  E_{k'\omega'}}\right\rangle\nonumber\\ &\,=\,&\int \frac{{\rm
    d}^3k}{(2\pi)^3}\frac{{\rm d}\omega}{2\pi}\,
\Im\left(\frac{1}{\omega\chi_{k\omega,\rm
    T}}\right){\cal S}_{\delta j}(k,\omega) .
\end{eqnarray}
The last equality uses the relation $\bs{\delta
  E_{k\omega}}\,=\,i\,\bs{\delta
  j_{k\omega}}/\left(\omega\chi_{k\omega,\rm T}\right)$,
$\chi_{k\omega,\rm T}$ denoting the transverse susceptibility in
$\omega-k$ space. It also introduces the power spectrum of current
density fluctuations, through $\langle \delta j_{k\omega}\delta
j_{k'\omega'}^\star\rangle\,=\,
(2\pi)^4\delta\left(\bs{k}-\bs{k'}\right)\delta(\omega-\omega') {\cal
  S}_{\delta j}(k,\omega)$. 

The transverse current density fluctuations are related to the
transverse magnetic modes through~\citep{2013PhPl...20e2113F}
\begin{equation}
\bs{\delta j_{k\omega}}\,=\,\frac{i}{4\pi}\bs{k}c\times\bs{\delta
    B_{k\omega}}\left[1+\frac{\vert\omega\vert^4}{(kc)^4}\right]\ .\label{eq:djdB}
\end{equation} 
One can safely neglect the last term in the brackets since $\vert
\omega\vert=\gamma_k$, the damping rate, and $\gamma_k\,\ll\,kc$ as
demonstrated further on.  Therefore the power spectra of current
fluctuations and magnetic turbulence are related through ${\cal
  S}_{\delta j}(k,\omega)\,=\,(k c/4\pi)^2{\cal S}_{\delta
  B}(k,\omega)$ and for magneto-static turbulence, ${\cal S}_{\delta
  B}(k,\omega)\,=\,2\pi\delta(\omega)\,{\cal S}_{\delta B}(k)$. One
thus finally arrives at
\begin{equation}
\frac{{\rm d}\left\langle\delta B^2\right\rangle}{{\rm
    d}t}\,=\,-2\int\frac{{\rm d}^3k}{(2\pi)^3}\,
\gamma_{k}\,{\cal S}_{\delta B}(k)\ ,
\end{equation}
with damping rate in $k-$space ($\gamma_k$ is counted as positive for
effective damping):
\begin{equation}
\gamma_{k}\,=\,-k^2c^2\Im\left(\frac{1}{4\pi\omega\chi_{k\omega,\rm
    T}}\right)_{\omega\rightarrow 0}\ .
\end{equation}

Therefore, the bulk of the calculation consists in evaluating the
non-linear susceptibility. For reference, assuming ballistic
trajectories and low frequencies $\omega\,\ll\,k c$, one has
\begin{equation}
4\pi\chi_{k\omega,\rm T}\,\simeq\,-i\frac{\pi}{4}\sum_{\alpha}\frac{4\pi
  q_\alpha^2}{\omega k}\,\int{\rm d} p\,p^2\frac{{\rm d}\faz}{{\rm
    d}p}\,\Theta\left[1-\left(\frac{\omega}{kv}\right)^2\right]\,+\,{\cal
  O}(\omega^0)\ ,
\end{equation}
where $\faz(p)$ represents the homogeneous part of the distribution
function of particles of species $\alpha$.  For a J\"uttner-Synge
distribution:
\begin{equation}
\faz(p)\,=\,\frac{n_\alpha\mu}{4\pi m^3 c^3 K_2(\mu)}e^{-\mu\gamma} \, 
\end{equation}
with $\mu=mc^2/(kT)$, $\gamma = \left[1 + p^2/(mc)^2\right]^{1/2}$ and
$n_\alpha$ the density of particles, one finds as
$\omega\,\rightarrow\,0$
\begin{equation}
4\pi\chi_{k\omega,\rm
  T}\,\simeq\,i\frac{\pi}{4}\sum_\alpha\frac{\omega_{\rm
    p,\alpha}^2}{\omega kc}\frac{1}{K_2(\mu)}\left(\frac{2}{\mu} +
\frac{2}{\mu^2}\right)e^{-\mu}\ ,
\end{equation}
in terms of the relativistic plasma frequency (squared) $\omega_{\rm
  p,\alpha}^2\,=\,4\pi n_\alpha q_\alpha^2 \mu/m$ which leads to the
ultra-relativistic ($\mu\,\rightarrow\,0$) linear damping rate 
\begin{equation}
\gamma_{k}\,\simeq\,\frac{4}{\pi}\frac{k^3c^3}{\omega_{\rm
    p}^2}\ ,
\end{equation}
with $\omega_{\rm p}^2\,=\,\sum_\alpha \omega_{\rm p,\alpha}^2$ the
relativistic plasma frequency of the global plasma. This result for
the linear Landau damping rate in a ultra-relativistic plasma matches
previous derivations, e.g. \citet{2001PhPl....8.1482B},
\citet{2008ApJ...674..378C} and \citet{2013PhPl...20j4502F}.

One can generalize very easily the above result to a power-law
distribution of particles with index $s$ and minimum Lorentz factor
$\gamma_{\rm min}$:
\begin{equation}
\faz(p)\,=\,\frac{n_\alpha\,\vert s-1\vert}{4\pi m^3c^3\gamma_{\rm
    min}}\left(\frac{\gamma}{\gamma_{\rm
    min}}\right)^{-s-2}\,\Theta(\gamma-\gamma_{\rm min})\ .
\end{equation}
One then infers in the ultra-relativistic limit $\gamma_{\rm
  min}\,\gg\,1$
\begin{equation}
\gamma_{k}\,=\,\frac{4}{\pi}\frac{k^3c^3}{\omega_{\rm
    p}^2}\frac{s}{\vert(s+2)(s-1)\vert}\ .
\end{equation}
The damping rate differs from the previous by a factor of order unity
only. In the following, the calculation of the non-linear damping rate
will be carried out for this power-law distribution function, since it
guarantees that there are no particle with Lorentz factor outside the
range of application of the approximation used (see further
below). Furthermore, one expects the distribution function in
astrophysical blast waves to follow such a power-law to a good
approximation; notably, Fermi acceleration at relativistic shock waves
predicts a spectral index $s\,\simeq\,2.3$ in the ultra-relativistic
limit for isotropic
scattering~\citep[e.g.][]{1998PhRvL..80.3911B,2000ApJ...542..235K,2001MNRAS.328..393A,2003ApJ...589L..73L,2005PhRvL..94k1102K}.

\subsection{Non-linear susceptibility}\label{sec:nls}
The current density fluctuations, from which one can extract the
susceptibility, are defined in terms of the fluctuating part of the
distribution function, as:
\begin{equation}
\bs{\delta j}\,=\,\sum_{\alpha}q_\alpha\int {\rm
  d}^3r\,\,\bs{v}\,\delta\fabolt \ .\label{eq:curr}
\end{equation}
The full distribution function is written $\fabolt\,=\, \faz(\bs{p},t)
+ \delta \fabolt$, with $\delta \fabolt$ the random inhomogenous part
and $\faz(\bs{p},t)$ the spatial average. Following
\citet{1969PhFl...12.1045W}, \citet{1970PhFl...13.2308W},
\citet{1975Ap&SS..38..125B} this fluctuating part is given by the
solution to the inhomogeneous part of the Boltzmann equation, and it
can be written in terms of a propagator ${\cal U_A}$ as:
\begin{equation}
\delta \fabolt\,=\,{\cal U_A}(t,t_0)\delta f_\alpha(\bs{r},\bs{p},t_0) - 
\int_{t_0}^{t}{\rm d}\tau\,\,{\cal U_A}(t,\tau)\,\bs{\delta{\cal F}}(\tau)
\cdot\frac{{\rm d}\faz(\bs{p},\tau)}{{\rm d}\bs{p}}\ .
\end{equation}
The random force operator is $\bs{\delta{\cal F}}(\tau)\,\equiv\,
q_\alpha\left[\bs{\delta E}(\bs{r},\tau) +\bs{v}\times\bs{\delta
    B}(\bs{r},\tau)/c\right]$. In the following, $\faz$ is assumed
isotropic in $p$; then the term associated to the magnetic Lorentz
force vanishes in the above expression.

The properties of ${\cal U_A}$ are described in details in the above
references and its relation to other propagators is discussed
in~\citet{1971PhFl...14.2239B}. For the sake of completeness, their
definitions are recalled in Appendix~\ref{app:propagators}.

In the following, the initial data will be written $\indat$ out of
brevity and clarity.  Going over to Fourier variables,
\begin{eqnarray}
\delta \fakbolt&\,=\,& \indat - \nonumber\\
&& q_\alpha\,\int {\rm d}^3r\int
\frac{{\rm
    d}^3k'}{(2\pi)^3}\,e^{-i\bs{k}\cdot\bs{r}}
\int_{t_0}^{t}{\rm d}\tau\,\,{\cal
  U_A}(t,\tau)\,e^{i\bs{k'}\cdot\bs{r}} \bs{\delta E_{k'}}(\tau)
\cdot \frac{\bs{v}}{v}\frac{{\rm d}}{{\rm d}
  p} \faz(p,t)\ .\nonumber\\\label{eq:fakbolt}
\end{eqnarray}
So far, the treatment has been exact; in particular, the separation of
$\fabolt$ into an average and a random part does not imply any
linearisation procedure.  The main approximation of the present work
is to approximate the full propagator ${\cal U_A}$ by the average
propagator $\overline{\cal U}$, which corresponds to the truncation to
the first term in a series expansion in powers of the fields, see
App.~\ref{app:propagators}, which summarizes the properties of these
propagators, and see most notably \citet{1966PhFl....9.1773D},
\citet{1969PhFl...12.1045W}, \citet{1970PhFl...13.2308W},
\citet{1971PhFl...14.2239B} and \citet{1975Ap&SS..38..125B}.  As
recalled in App.~\ref{app:propagators}, higher order terms are
suppressed relative to this first order correction by powers of
$c\tau_{\rm c}/r_{\rm g}$, with $\tau_{\rm c}$ the correlation time of
the electromagnetic fluctuations, $r_{\rm g}$ the typical gyroradius
of the particles in the turbulence, defined with respect to $\langle
\delta B^2\rangle^{1/2}$. The present work thus makes the explicit
  assumption that $r_{\rm g}\,>\,c\tau_{\rm c}$.

In relativistic blast waves, the typical Lorentz factor of a particle
downstream of a relativistic shock wave of Lorentz factor $\gamma_{\rm
  sh}$ is $\gamma_{\rm sh}$ for a pair shock, or $\gamma_{\rm sh}$
(resp. $\gamma_{\rm sh}m_{\rm i}/m_e$) for the ion (resp. electron)
population in an electron-ion
shock~\citep[e.g.][]{2008ApJ...673L..39S,2008ApJ...682L...5S}. In the
following, this Lorentz factor is denoted $\gamma_{\rm min}$.  One
then derives the typical ratio $r_{\rm g}/c\tau_{\rm c}$ for a
particle of Lorentz factor $\gamma$:
\begin{equation}
\frac{r_{\rm g}}{c\tau_{\rm c}}\,\simeq\,
\epsilon_B^{-1/2}\frac{k_{\rm max}c}{\omega_{\rm
    p}}\frac{\gamma}{\gamma_{\rm min}}\ ,
\end{equation}
The typical scale of Weibel turbulence is $c\tau_{\rm c}\,=\,k_{\rm
  max}^{-1}\,\sim\, {\cal K}\,c/\omega_{\rm p}$ with ${\cal
  K}\,\simeq\,10$ close to the shock
front~\citep{2008ApJ...673L..39S,2008ApJ...674..378C,2009ApJ...693L.127K,2013ApJ...771...54S}.
Given that $\epsilon_B\,\lesssim\,10^{-2}$, this indicates that
typically, $r_{\rm g}\,\gtrsim\,c\tau_{\rm c}$, possibly $r_{\rm
  g}\,\gg\,c\tau_{\rm c}$, depending on $\gamma/\gamma_{\rm min}$ and
$\epsilon_B$. The expansion used here should therefore be a good
approximation away from the shock front, where
$\epsilon_B\,\lesssim\,10^{-2}$.

It is instructive to rewrite the above expansion parameter in terms of
the ratio of fluctuating to mean quantities. In particular, using
Eq.~(\ref{eq:djdB}), which relates the current fluctuations to the
magnetic fluctuations, one can show that, in orders of magnitude,
$\delta n/n \,\sim\, k_{\rm max}c\, \delta B_{k_{\rm max}}/(n
ec)\,\sim\,\epsilon_B^{1/2} k_{\rm max}c/\omega_{\rm p}$, with $\delta
n$ the density of current-carrying fluctuations. Therefore the above
hierarchy $r_{\rm g}/(c\tau_{\rm c})\,>\,1$ at $\gamma_{\rm min}$ also
implies $\delta n\,<\,n$, i.e. small fluctuations; note that the
former constraint $r_{\rm g}/(c\tau_{\rm c})\,>\,1$ is more stringent
than the latter $\delta n\,<\,n$, because ${\cal K}\,\gtrsim\,10$.

Since $\bs{\delta E_{k'}(\tau)}$ depends solely on time, it commutes
with ${\cal U_A}$ (see
App.~\ref{app:propagators}). Equation~(\ref{eq:fakbolt}) can then be
approximated as
\begin{eqnarray}
&&\delta \fakbolt\,=\,\indat - \nonumber\\ &&\quad\quad q_\alpha\,\int
  {\rm d}^3r\int \frac{{\rm
      d}^3k'}{(2\pi)^3}\,e^{-i\bs{k}\cdot\bs{r}+i\bs{k'}\cdot\bs{r}}
  \int_{t_0}^{t}{\rm d}\tau\,\bs{\delta
    E_{k'}(\tau)}\cdot\left\langle\,e^{i\bs{k'}\cdot\bs{\Delta
      r_s}(\tau)} \frac{\bs{v_s}(\tau)}{v_s}\frac{{\rm
      d}\faz(p,\tau)}{{\rm d} p}\right\rangle\ .\nonumber\\
\end{eqnarray}
The quantities $\bs{r_s}(\tau)$ and $\bs{v_s}(\tau)$ represent the
exact orbits of the particles in the fluctuating fields at time $\tau$
with boundary conditions $\bs{r_s}(t)=\bs{r}$, $\bs{v_s}(t)=\bs{v}$;
furthermore, $\bs{\Delta r_s}(\tau)=\bs{r_s}(\tau)-\bs{r}$. The
average over the exact orbits will be calculated further on in the
limit of a magneto-static turbulence. In this limit, $v_s$ and ${\rm
  d}\faz(p)/{\rm d}p$ are constant in time; thus, $v_s=v$ in
particular and these terms can be extracted from the average. Of
course, dissipation is accompanied by a transfer of energy from the
fields to the particles. This, however, takes place on a timescale
$\sim\,\gamma_k^{-1}$ much larger than the scattering time of the
particles, $t_{\rm s}$, so that on this latter timescale, energy flow
can indeed be neglected. Furthermore, in relativistic blast waves, the
turbulence energy density contains much less energy than the
particles, $\epsilon_B\,\ll\,\epsilon_e\,\sim\,0.1$ (see above
references).

As a result of spatial homogeneity, the average does not depend on
$\bs{r}$; it only depends on $\bs{v}$ and $t-\tau$. Therefore
\begin{equation}
\delta \fakbolt\,=\,\indat - q_\alpha\,\int_{t_0}^{t}{\rm
  d}\tau\,\bs{\delta
  E_{k}(\tau)}\cdot\left\langle\,e^{i\bs{k}\cdot\bs{\Delta r_s}(\tau)}
\bs{v_s}(\tau)\right\rangle\frac{1}{v}\frac{{\rm d}\faz(p)}{{\rm d}p}
\ .\nonumber\\
\end{equation}
Similarly, using the fact that the above expression is written as a
convolution in time, the Laplace-Fourier transform of the fluctuating
part of the distribution function ends up being
\begin{equation}
\delta \faksbolt\,=\,\indat -  
q_\alpha\,{\rm L}_{\rm L}\left\langle\,e^{i\bs{k}\cdot\bs{\Delta r_s}(\tau)}
\bs{v_s}(\tau)\right\rangle\cdot\bs{\delta
  E_{k,\omega}}\frac{1}{v}\frac{{\rm d}\faz(p)}{{\rm d} p}\ ,
\end{equation}
with ${\rm L_{\rm L}}$ the Laplace transform operator; $\bs{\delta
  E_{k,\omega}}$ represents the Fourier-Laplace transform of the
fluctuating electric field. Omitting the initial data, from
Eq.~(\ref{eq:curr}) and $\bs{\delta j_{k\omega}}\,\equiv\,
-i\omega\chi_{k\omega}\,\bs{\delta E_{k\omega}}$, one then
extracts the non-linear susceptibility
\begin{equation}
\chi_{k\omega\,\,ij}\,=\,-\sum_{\alpha} \frac{q_\alpha^2}{-i\omega}\int{\rm d}^3p
\,\frac{{\rm d}\faz}{{\rm d}p}\,\frac{v_i}{v}\,{\rm L}_{\rm
  L}\left\langle\,e^{i\bs{k}\cdot\bs{\Delta r_s}(\tau)}
v_{s\,j}(\tau)\right\rangle\ .\label{eq:chi}
\end{equation}
One is particularly interested in the transverse susceptibility, since
the turbulence is assumed magnetostatic:
\begin{equation}
\chi_{k\omega,\rm
  T}\,\equiv\,\frac{1}{2}\left(\delta^{ij}-\frac{k^ik^j}{k^2}\right)\chi_{ij}\ .
\end{equation}

\section{Analytical approximations and results}\label{sec:nld-calc}
\subsection{Non-linear susceptibility}
In order to calculate the first-order non-linear correction to the
above damping rate, one now needs to evaluate the average over the
exact orbits in Eq.~(\ref{eq:chi}). This is done in
Appendix~\ref{app:correlators}.

Note that Appendix~\ref{app:correlators} assumes explicitly that the
magnetic field behaves as white noise, with zero average, with a
correlation time $\tau_{\rm c}$ assumed to be smaller than the
scattering timescale of the particles $t_{\rm s}$. Physically, this
corresponds to the transport of particles in a small-scale turbulence,
i.e. to the same approximation as above, $c\tau_{\rm c}\,<\,r_{\rm
  g}$, with $r_{\rm g}$ the typical gyroradius of the
particle defined in terms of the rms magnetic field.

The transport of particles in small-scale magneto-static turbulence is
well known, see in particular \citet{2011A&A...532A..68P} for a recent
study. In this configuration, one can work out exactly the correlators
that appear in Eq.~(\ref{eq:chi}), see App.~\ref{app:correlators}.
One thus derives
\begin{eqnarray}
4\pi\chi_{k\omega,\rm T}&\,\simeq\,&-\frac{1}{2}\sum_\alpha \frac{4\pi
  q_\alpha^2}{-i\omega}2\pi\int{\rm d}p{\rm d}\mu \, p^2 v \frac{{\rm
    d}\faz}{{\rm d}p}\, \int_0^{+\infty}{\rm d} t\, \nonumber\\
&&
\exp\Biggl\{
+i\omega t-ikvt_{\rm s}\mu(1-C_1)-\frac{1}{2}k^2v^2t_{\rm
  s}^2\mu^2\left[-\frac{1}{3}+C_1-C_2+\frac{1}{3}C_3\right]\nonumber\\
&&\quad\quad-\frac{1}{3}k^2v^2t_{\rm s}^2\left[\frac{t}{t_{\rm
    s}}-\frac{4}{3}+\frac{3}{2}C_1- \frac{1}{6}C_3\right]\Biggr\}\nonumber\\
&&\quad\quad\times \left[\left(1 + \frac{ikv\mu t_{\rm s}}{2}\right)C_1 -
  ikvt_{\rm s}\mu C_2 + \frac{ikv\mu t_{\rm
      s}}{2}C_3\right]\nonumber\\ \ ,\label{eq:chifull}
\end{eqnarray}
using the short-hand notation: $C_p\,\equiv\,\exp\left(-p\,t/t_{\rm
  s}\right)$.  The variable $\mu$ is defined as the cosine of the
angle between the wave vector $\bs{k}$ and $\bs{p}$. The time integral
explicits the Laplace transform over the correlation function. The
above expression represents the main result of the present paper.

One notes that $k v t_{\rm s}\,\simeq\,t_{\rm s}/\tau_{\rm c}\,>\,1$,
since $k\,\simeq\,1/(c\tau_{\rm c})$ for a magneto-static turbulence
on scale $k^{-1}$. One can thus approximate the above result as
follows. First of all, one notes that the exponential contained in
Eq.~(\ref{eq:chifull}) is cut off at large times, due to the
decorrelation of the particle trajectories. Introducing the following
large parameter:
\begin{equation}
\kappa\,\equiv\, k v t_{\rm s}\ ,
\end{equation}
which explicitly depends on particle momenta through $v$ and $t_{\rm
  s}$, expanding the terms in the exponential in the limit
$t\,\ll\,t_{\rm s}$, one obtains Eq.~(\ref{eq:corrsmallt}), which
reveals that the cut-off becomes prominent whenever $\kappa^2
t^3/t_{\rm s}^3\,<\,1$, i.e. $t\,<\,\kappa^{-2/3}t_{\rm s}$.  Since
$\kappa\,>\,1$, this justifies the approximation of the above integral
in the small-time limit $t\,\ll\,t_{\rm s}$:
\begin{eqnarray}
4\pi\chi_{k\omega,\rm T}&\,\simeq\,&-\frac{1}{2}\sum_\alpha \frac{4\pi
  q_\alpha^2}{-i\omega}2\pi\int{\rm d}p{\rm d}\mu \, p^2 v \frac{{\rm
    d}\faz}{{\rm d}p}\,\nonumber\\
&& \quad\int_0^{+\infty}{\rm d} t\, \exp\left[
+i\omega t-i \kappa\mu t/t_{\rm s}-\frac{1}{6}\left(1-\mu^2\right)\kappa^2t^3/t_{\rm
  s}^3\right]\nonumber\\
&&\quad\quad\times \left[\left(1 + \frac{i\kappa\mu}{2}\right)C_1(t) -
  i\kappa\mu C_2(t) + \frac{i\kappa \mu}{2}C_3(t)\right]
\end{eqnarray}
One can check that in the limit $t_{\rm s}\,\rightarrow\,+\infty$, one
recovers the linear transverse susceptibility, as expected. This
integral is of the Airy type. It can be written and further
approximated by
\begin{equation}
4\pi\chi_{k\omega,\rm T}\,=\,-i\pi\sum_\alpha \frac{4\pi
  q_\alpha^2}{\omega}\int{\rm d}p{\rm d}\mu \, p^2 v \frac{{\rm
    d}\faz}{{\rm d}p}\,\left[\left(1+\frac{i\kappa}{2}\right)I_1 -
i\kappa I_2 + \frac{i\kappa}{2}I_3\right]\left(1-\mu^2\right)\ ,
\end{equation}
with
\begin{eqnarray}
I_p&\,\equiv\,&t_{\rm s}\int_0^{+\infty}{\rm d}\hat
t\,\exp\left[i\left(\omega t_{\rm s}-\kappa\mu +ip\right)\hat t
  -\frac{1-\mu^2}{6}\kappa^2\hat t^3\right]\nonumber\\
&\,\approx\,& \frac{t_{\rm s}}{-i\omega t_{\rm s}+i\kappa\mu + p
    +\left[\frac{1-\mu^2}{6}\kappa^2\right]^{1/3}}\ .
\end{eqnarray}
Finally, one can work out the integral over $\mu$ after dropping the
slow dependence on $(1-\mu^2)^{1/3}$ in the denominators, i.e. making
the substitution $(1-\mu^2)^{1/3}\,\sim\,1$; this leads to
\begin{eqnarray}
4\pi\chi_{k\omega,\rm T}&\,\simeq\,&-\pi\sum_\alpha \frac{4\pi
  q_\alpha^2}{\omega k}\int{\rm d}p{\rm d}\mu \, p^2 v \frac{{\rm
    d}\faz}{{\rm
    d}p}\,\Biggl\{(1-m_1)^2\ln\left(-\frac{1-m_1}{1+m_1}\right) -
2m_1\nonumber\\
&&\quad\quad +i\kappa\left[\frac{1}{2}m_1(1-m_1^2)\ln\left(-\frac{1-m_1}{1+m_1}\right)
  - m_2(1-m_2^2)\ln\Biggl(-\frac{1-m_2}{1+m_2}\right)\nonumber\\
&&\quad\quad\quad\quad
  +\frac{1}{2}m_3(1-m_3)^2\ln\left(-\frac{1-m_3}{1+m_3}\right) - m_1^2
  + 2m_2^2 - m_3^2\Biggr]\Biggr\}\label{eq:chiapp}
\end{eqnarray}
with the short-hand notation:
\begin{equation}
m_p\,\equiv\, \frac{\omega}{k v}\,+\,i\left(p\kappa^{-1} +
6^{-1/3}\kappa^{-2/3}\right)\ .
\end{equation}
In the limit $t_{\rm s}\,\rightarrow\,+\infty$,
$\kappa\,\rightarrow\,+\infty$ and $m_p\,\rightarrow\,\omega/(k v)$;
Eq.~(\ref{eq:chiapp}) then reduces to the standard (linear) expression
for the transverse susceptibility of a relativistically hot
plasma. Note that in the limit $\omega\,\rightarrow\,0$, one can
expand $\omega\chi_{k\omega,\rm T}$ to lowest order in negative powers
of $\kappa$, yielding:
\begin{eqnarray}
\left.4\pi\omega\chi_{k\omega,\rm T}\right\vert_{\omega\rightarrow
  0}&\,\approx\,& -\pi\sum_\alpha \frac{4\pi q_\alpha^2}{
  k}\int{\rm d}p{\rm d}\mu \, p^2 v \frac{{\rm d}\faz}{{\rm
    d}p}\,\nonumber\\
&&\quad\quad\times\left[i\pi - 2^{5/3}3^{-1/3}i \kappa^{-2/3} + i
  6^{-2/3}\pi\kappa^{-4/3}+\ldots\right]\ .
\end{eqnarray}
The $i\pi$ term within the brackets corresponds to the linear result;
the lowest order term in $\kappa^{-2/3}$ indicates that the non-linear
effects  tend to increase the damping rate. This will be confirmed in
the full calculation below.

\subsection{Non-linear damping rate vs $k$}
In order to evaluate the above integrals and recast them in a proper
context, one needs to explicit the dependence of $\kappa$ on
wavenumber and momenta; since $t_{\rm s}\,\propto\, r_{\rm g}^2$ and
$\kappa = k v t_{\rm s}$, $\kappa\,\propto\,\gamma^2$ of course. One
now assumes that the power spectrum of magnetic turbulence in Fourier
space takes on a power-law shape and peaks at some maximum wavenumber
$k_{\rm max}$, ${\cal S}_{\delta B}(k)\,\propto\, (k/k_{\rm
  max})^{n_B}$, with $n_B\,>\,-2$ to guarantee $\tau_{\rm
  c}\,\sim\,1/(k_{\rm max}c)$. Linear theory predicts a damping rate
$\gamma_k \,\propto\, k^3$, indicating that damping is much faster on
the smaller spatial scales, as expected. Therefore, behind the shock
front, the turbulent power spectrum is built up at some initial time,
then gets eroded as time goes on, smaller scales being dissipated
first. In short, the maximum wavenumber, at which there remains net
power, becomes time-dependent. At a given time, one should therefore
evaluate the damping rate of the turbulence at the (time-dependent)
maximum wavenumber $k_{\rm max}(t)$, since modes with smaller
wavenumbers will be damped on much longer timescales. One can relate
the magnetic power at time $t$ to the initial power through
$\left\langle\delta B(t)^2\right\rangle\,\simeq\,\left\langle\delta
B(0)^2\right\rangle \left[k_{\rm max}(t)/k_{\rm
    max}(0)\right]^{n_B+3}$.  In this way, recalling that $t_{\rm
  s}\,=\,(3/2)\gamma^2m^2c^2/\left(\tau_{\rm c}e^2\langle\delta
B^2\rangle\right)$ (see App.~\ref{app:correlators}), with $\tau_{\rm
  c}\,\simeq\,\left[k_{\rm max}(t)c\right]^{-1}$, one finds
\begin{equation}
\kappa\,=\,\kappa_0 \left[\frac{k_{\rm max}(t)}{k_{\rm
    max}(0)}\right]^{-n_B-1}\left(\frac{\gamma}{\gamma_{\rm min}}\right)^2
\ ,
\end{equation}
with $\kappa_0$ the value of $\kappa$ at $t=0$, at $k_{\rm max}(0)$
and at $\gamma_{\rm min}$.

Interestingly, depending on the shape of the power spectrum of
magnetic fluctuations, one can find situations where $\kappa$
increases or decreases as a function of $k$; in the limiting case
$n_B=-1$, $\kappa$ becomes independent of the (time-dependent) maximum
wavenumber behind the relativistic shock wave, meaning that $\kappa$
does not depend on time (since injection through the shock) or,
equivalently, on distance to the shock front in the downstream plasma
rest frame (the shock front moving away at velocity $c/3$ in that
frame). However, if $n_B<-1$, $\kappa$ decreases with decreasing
wavenumber, because erosion leaves enough power at low $k-$modes,
while the effective coherence length increases, thereby leading to the
eventual trapping of particles, $\kappa\,<\,1$. Since the present
calculations rely on the approximation $\kappa\,>\,1$, the following
assumes $n_B>-1$. The PIC simulations of \citet{2008ApJ...674..378C}
further suggest that indeed $n_B$ is closer to zero, although this
ignores the influence of high-energy particles on the turbulence, as
discussed in \citet{2009ApJ...693L.127K} and
\citet{2009ApJ...696.2269M}.

\begin{figure}
  \centerline{\includegraphics{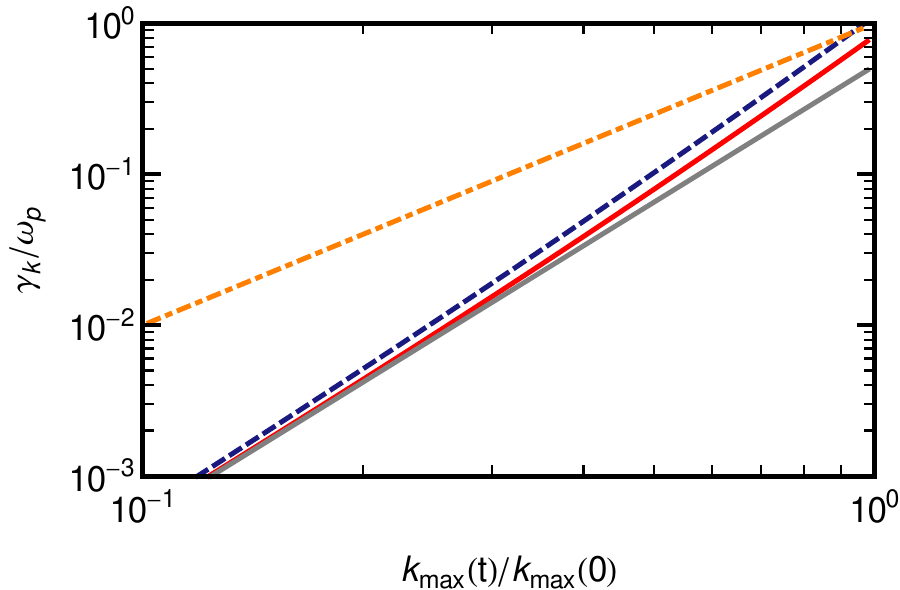}}
  \caption{Damping rate vs wavenumber: in thick gray, the linear
    calculation; in thick red, the damping rate at first non-linear
    order, assuming $\kappa_0\,=\,1$ and $n_B\,=\,0$, calculated
    through Eq.~(\ref{eq:chifull}); in dashed blue, the same damping
    rate calculated with the approximation Eq.~(\ref{eq:chiapp}); in
    dash-dotted orange, the scattering frequency $t_{\rm s}^{-1}$ in
    units of $\omega_{\rm p}$ at $\gamma_{\rm min}$.}
\label{fig:f1}
\end{figure}

\begin{figure}
  \centerline{\includegraphics{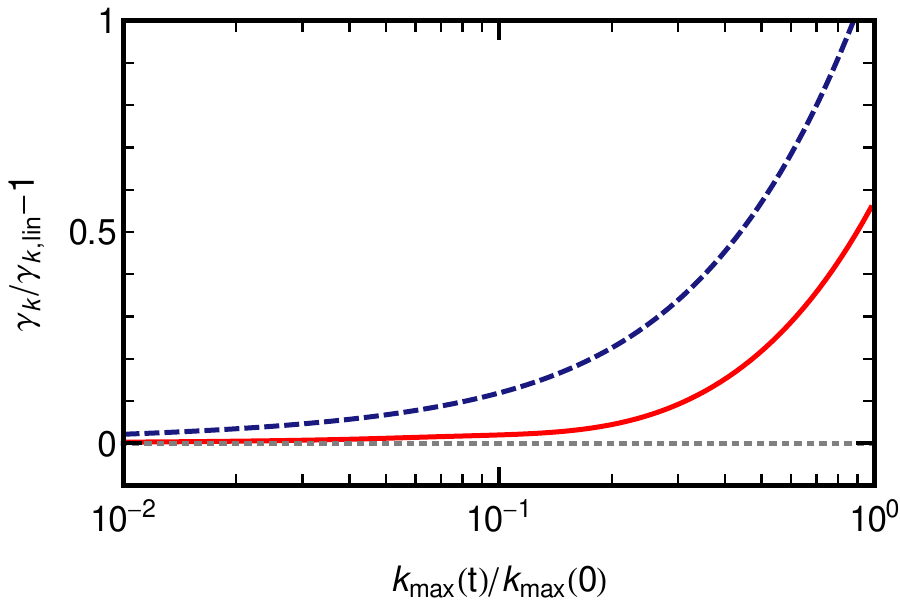}}
  \caption{Evaluation of the non-linear correction to the linear
    damping rate ($\gamma_{k,\rm lin}$) vs wavenumber: in thick red,
    the correction calculated using Eq.~(\ref{eq:chifull}); in dashed
    blue, the correction calculated with the approximation
    Eq.~(\ref{eq:chiapp}). As Fig.~\ref{fig:f1}, this figure assumes
    $\kappa_0\,=\,1$ and $n_B\,=\,-1$.}
\label{fig:f2}
\end{figure}

Figures~\ref{fig:f1} and \ref{fig:f2} show a numerical evaluation of
the damping rate $\gamma_k$, obtained through a full calculation of
Eq.~(\ref{eq:chifull}) and of its approximation Eq.~(\ref{eq:chiapp}),
as a function of the time-dependent $k_{\rm max}$, assuming
$n_B\,=\,0$ and $\kappa_0\,=\,1$.

Figure~\ref{fig:f1} also shows the evolution of the scattering
frequency $t_{\rm s}^{-1}$ vs wavenumber: this allows to verify that,
at all $k_{\rm max}$, one has $k^{-1}\,<\,t_{\rm
  s}\,<\,\gamma_k^{-1}$, which validates the assumptions inherent to
the present approach.

These figures show that the non-linear calculation modifies the linear
calculation by a factor of order unity at $k_{\rm max}(0)$, then
converges to the linear calculation at smaller $k_{\rm max}$; indeed,
$n_B\,=\,0$ implies $\kappa\,\propto\, k_{\rm max}^{-1}$: $\kappa$
increases with decreasing values of $k_{\rm max}$ at a same
$\gamma_{\rm min}$, therefore the importance of non-linear effects,
which is quantified by inverse powers of $\kappa$, becomes weaker as
$k_{\rm max}$ decreases. As mentioned above, for $n_B\,=\,-1$, one
would find a correction at all $k_{\rm max}$ equal to the correction
calculated at $k_{\rm max}(0)$.

These calculations also indicate that the non-linear terms
systematically lead to an increased damping rate, although the
correction is modest. This goes contrary to the discussion in
\citet{2008ApJ...674..378C}, which conjectured that the deflection of
particles by magnetic turbulence might lead to a weaker damping rate.

\section{Discussion and conclusions}\label{sec:disc}
The present work studies the damping rate of the micro-turbulence
which has been excited through e.g. Weibel/filamentation instabilities
in the precursor of a weakly magnetized relativistic collisionless
shock wave then transmitted downstream. As mentioned in
Sec.~\ref{sec:intr}, such calculations are directly relevant to the
physics of collisionless shock waves, but also to high energy
astrophysics, since the damping of the turbulence governs the strength
of the magnetic field in which electrons radiate (and therefore the
frequency at which they radiate the bulk of their energy).

In the standard afterglow model for gamma-ray burst, the canonical
value for the equipartition fraction of magnetic energy density in the
blast is taken as $\epsilon_B\,\sim\,10^{-2}$, on the basis on
afterglow observations in various wavebands, see
e.g. \citet{1997ApJ...485L...5W}, \citet{1999ApJ...523..177W} for
early determinations, and \citet{2001ApJ...560L..49P} for a
compilation of results, which however reveals a large scatter in this
parameter. Such a value would fit nicely the results of PIC
simulations in the absence of dissipation, since these simulations
find $\epsilon_B\,\sim\,10^{-2}$ immediately downstream of the blast;
the fact that $\epsilon_B$ remains that large up to the long time
scales on which electrons can radiate gives rise to the notorious
problem of the origin of these magnetic fields.

Recent detections of gamma-ray burst afterglows at high-energy
$>100\,$MeV may have shed a new light on this issue. If this
high-energy emission indeed corresponds to the synchrotron afterglow,
e.g.~\citep{2009MNRAS.400L..75K,2010MNRAS.409..226K,2010MNRAS.403..926G},
these detections offer another observational constraint to pin down
$\epsilon_B$ beyond the degeneracies inherent to most of the previous
studies, see the discussion in \citet{2013MNRAS.435.3009L}. Then one
derives low values of $\epsilon_B$, well below the canonical one,
which may be interpreted as the partial dissipation of the
Weibel-generated turbulence, as described here
\citep{2013MNRAS.428..845L}; in particular, assuming a power-law decay
$\epsilon_B\,\propto\,(t\omega_{\rm p})^{\alpha_t}$ as a function of
comoving time, one derives $-0.5\,\lesssim\,\alpha_t\,\lesssim\,-0.4$
from a handful of gamma-ray burst afterglows seen in radio, optical,
X-ray and at high energy \citep{2013MNRAS.435.3009L}, i.e. a net
dissipation. The decay of Weibel turbulence behind relativistic shock
waves thus appears as a key ingredient in describing accurately the
light curves of these extreme astronomical phenomena.

The present work presents a calculation of the damping rate to the
first non-linear order, by computing the effects of particle diffusion
in the micro-turbulence. As mentioned in Sec.~\ref{sec:intr}, one
interest of such a calculation is to study the dependence of this
correction on the power spectrum of magnetic fluctuations, which at
present cannot be reconstructed with confidence by PIC simulations. An
exact calculation is possible, thanks to the small-scale and
magneto-static nature of the turbulence, which allows for an explicit
calculation of the trajectory correlators which determine the amount
of resonance broadening. As discussed in the main text, this work
assumes that the particles are not trapped in the micro-turbulence,
i.e. the scattering length is assumed larger than the coherence length
of the magnetic fluctuations.

The overall influence of non-linear terms is found to be of order
unity at the maximum wavenumber, and to decrease with decreasing
wavenumbers $k$, provided the three-dimensional power spectrum of
magnetic fluctuations ${\cal S}_{\delta B}\,\propto\,k^{n_B}$ has an
index $n_B\,>\,-1$. In this case, indeed, the ratio of the particle
scattering timescale $t_{\rm s}$ to the coherence time $\tau_{\rm c}$
of the magnetic fluctuations increases, therefore the particle
trajectories become more and more ballistic as $k$ decreases and one
recovers the linear result in the small $k$ limit. The results
obtained also indicate that the non-linear correction systematically
increases the damping rate.

The present results would suggest that the damping rate does follow
roughly the scaling $\gamma_k\,\propto\,k^3$ predicted by linear
theory, however one cannot exclude at present that the power spectrum
of magnetic fluctuations is such -- i.e. $n_B\,<\,-1$ -- that effects
associated to particle trapping become more and more prominent as
dissipation progresses (meaning, as the time-dependent maximum
wavenumber decreases). Actually, if $\gamma_k\,\propto\,k^{\alpha_k}$,
one can relate the decay exponent $\alpha_t$ to the power spectrum
index $n_B$ and
$\alpha_k$~\citep{2008ApJ...674..378C,2013MNRAS.428..845L}:
\begin{equation}
\alpha_t\,=\,-\frac{3+n_B}{\alpha_k}\ .
\end{equation}
Then, $\alpha_k\,\simeq\,3$ with a value $\alpha_t\,\sim\,-0.5$ as
suggested by observations~\citep{2013MNRAS.435.3009L} would imply
$n_B\,\sim\,-1.5$, in which case the ratio $t_{\rm s}/\tau_{\rm
  c}\,\propto\,k^{-n_B-1}$ would decrease with decreasing $k$: i.e.,
the non-linear effects would become more prominent as dissipation
progresses.

The present calculations cannot address the situation in which
particles are effectively trapped and other theoretical tools are
needed to probe this regime and to make the connection with
observations. Further PIC simulations, extended in time and
dimensionality, would also provide useful guidance to better
characterize the scaling of $\gamma_k$ vs $k$. Finally, dedicated PIC
simulations with an artificially set-up power spectrum might be used
to probe the regime $n_B\,<\,-1$ in which trapping is expected to
become more effective as dissipation progresses. \bigskip\bigskip

It is a pleasure to thank Laurent Gremillet and Guy Pelletier for very
valuable suggestions and discussions. This work has been financially
supported by the ``Programme National Hautes Energies'' (PNHE) of the
CNRS.

\appendix
\section{}\label{app:propagators}
Separating the average and random parts as usual, as described in
Sec.~\ref{sec:nls}, one finds that the fluctuating part of the
distribution function obeys:
\begin{equation}
\left[\frac{\partial}{\partial t} + {\cal L}\right]\fabolt - \left\langle\delta{\cal
L}\fabolt\right\rangle \,=\, - \delta {\cal L} \faz
\end{equation}
with:
\begin{equation}
{\cal L}\,=\,\overline{\cal L} + \delta{\cal L},\quad
\overline{\cal L}\,=\, \bs{v}\cdot\frac{\partial}{\partial\bs{r}},\quad
\delta{\cal L}\,=\, q\left(\bs{\delta
  E} + \frac{\bs{v}}{c}\times\bs{\delta B}\right)\cdot\frac{\partial}{\partial
  \bs{p}}
\end{equation}
Following~\citet{1969PhFl...12.1045W} and \citet{1970PhFl...13.2308W},
one introduces the averaging operator ${\cal A}$, which takes the
average over the statistical realization of the fluctuations of all
quantities to its right, i.e. $\psi_1(\bs{r},\bs{p},t){\cal
  A}\psi_2(\bs{r},\bs{p},t)\ldots\psi_n(\bs{r},\bs{p},t)\,=\,
\psi_1\langle\psi_2\ldots\psi_n\rangle$. One then defines the
following propagators:
\begin{eqnarray}
{\cal U}:&&\quad\left[\frac{\partial}{\partial t} + {\cal L}\right]{\cal
  U}(t,t_0)\,=\,0,\quad {\cal U}(t_0,t_0)\,=\,1 \nonumber\\
{\cal U_A}:&& \quad\left[\frac{\partial}{\partial t} + \left(1-{\cal A}\right){\cal L}\right]{\cal
  U_A}(t,t_0)\,=\,0,\quad {\cal U_A}(t_0,t_0)\,=\,1 \nonumber\\
{\cal \overline U}:&& \quad\left[\frac{\partial}{\partial t} + \overline{\cal L}\right]\overline{\cal
  U}(t,t_0)\,=\,-\left\langle\delta{\cal L}\,{\cal U}\right\rangle,\quad \overline{\cal U}(t_0,t_0)\,=\,1 \ .
\end{eqnarray}

${\cal U}(t,t_0)$ of course represents the full propagator of the
Vlasov equation; acting on a function $\psi(\bs{r},\bs{p},t)$, it
propagates it backward in time, i.e. ${\cal
  U}(t,t_0)\psi(\bs{r},\bs{p},t)\,=\,
\psi(\bs{r_s}(t_0),\bs{p_s}(t_0),t_0)$ with $\bs{r_s}(t_0)$,
$\bs{p_s}(t_0)$ the solutions of the characteristic equations for the
trajectories, such that $\bs{r_s}(t)=\bs{r}$ and $\bs{p_s}(t)=\bs{p}$.
The merit of the propagator ${\cal U_A}$ is to provide an explicit
solution for $\delta \fabolt$ in terms of its initial data and the
average distribution function $\faz$.  

The various propagators ${\cal U}$, ${\cal U_A}$ and $\overline{\cal U}$
are related through series expansions in powers of ${\cal
  L}$~\citep{1971PhFl...14.2239B}.  In order to obtain a tractable
expression for ${\cal U_A}$, one generally truncates such series to
the lowest order, which leads to ${\cal U_A}\,\simeq\,\overline{\cal
  U}$~\citep{1966PhFl....9.1773D,1969PhFl...12.1045W,1970PhFl...13.2308W}. Explicitly,
one finds to the next-to-leading order~\citep{1971PhFl...14.2239B}:
\begin{equation}
{\cal U}_A(t,t_0)\,=\,\overline{\cal U}(t,t_0) - \int_{t_0}^{t}{\rm
  d}\tau_1\,\,\overline{\cal U}(t,\tau_1)\left(1-{\cal A}\right){\cal
  L}(\tau_1)\overline{U}(\tau_1,t_0)\,+\ldots
\end{equation}
The magnitude of the next-to-leading order term relatively to the
first order term is $\tau_{\rm c}e\langle\delta
B^2\rangle^{1/2}/(\gamma m c)$, with $\gamma$ the Lorentz factor of
the particle.

 The propagator $\overline{\cal U}$ acts on a function $\psi$ by propagating
it backwards in time and taking the statistical average over the exact
orbits:
\begin{equation}
\overline{\cal U}(t,t_0)\psi(\bs{r},\bs{p},t)\,=\,\left\langle
\psi\left[\bs{r_s}(t_0),\bs{p_s}(t_0),t_0\right]\right\rangle\ ,
\end{equation}
with the boundary conditions $\bs{r_s}(t)=\bs{r}$ and
$\bs{p_s}(t)=\bs{p}$. Consequently, $\overline{\cal U}$ commutes with
quantities that depend solely on time.

\section{}\label{app:correlators}
This Appendix calculates the correlators over the characteristic
trajectories in the turbulence, which enter the expression
Eq.~(\ref{eq:chi}) for the non-linear susceptibility. All throughout
this section, the index $_s$ for the characteristic trajectories is
dropped, for clarity.

The particle suffers pitch-angle scattering in a magnetostatic
turbulence. As discussed in \citet{2011A&A...532A..68P}, a convenient
way to calculate the transport coefficients is to write the time
evolution of its velocity as a time-ordered product of an
exponentiated Liouville operator:
\begin{equation}
\bs{v}(t')\,=\,{\cal  T}\,\exp\left[-\int_{t'}^{t}{\rm d}\tau_1
  \widehat{\delta \Omega}\right]\bs{v}(t)\ ,\label{eq:vtordered}
\end{equation}
where a minus sign has been introduced in order to compute quantities
at time $t'<t$ as a function of quantities at time $t$. The rotation
operator $\widehat{\delta\Omega}$ is defined as
\begin{equation}
\widehat{\delta\Omega}\,=\,\delta\Omega^a\widehat L_a,\quad
\delta\Omega^a\,=\,\frac{e\delta B^a}{\gamma m c}
\end{equation}
with $\gamma$ the Lorentz factor of the particle, $a=1,2,3$, $\widehat
L_a$ a generator of the rotation group, with matrix components:
$\left.\widehat L_a\right._k\mbox{}^l\,=\,\epsilon_{ak}\mbox{}^l$
($\epsilon_{akl}$ denotes the Levi-Civita
symbol). Equation~(\ref{eq:vtordered}) solves the equation of motion
of the particle.

The rotation operator is assumed to behave as isotropic white noise
with correlation time $\tau_{\rm c}$:
\begin{equation}
\left\langle\delta\Omega^a(\tau_1)\right\rangle\,=\,0,\quad
\left\langle\delta\Omega^a(\tau_1)\delta\Omega^b(\tau_2)\right\rangle\,=\,
\frac{2}{3}\tau_{\rm
  c}\delta(\tau_1-\tau_2)\delta\Omega^2\delta^{ab}\ ,
\end{equation}
with $\delta\Omega^2\,=\,\left[e\langle\delta B^2\rangle^{1/2}/(\gamma
  m c)\right]^2$.  A useful identity is: $\delta ^{ab}\widehat
L_a\widehat L_b=-2 \widehat I$, which implies
$\left\langle\widehat{\delta\Omega}(\tau_1)\widehat{\delta\Omega}(\tau_2)\right\rangle
\,=\,-4\delta\Omega^2\tau_{\rm c}\delta(\tau_1-\tau_2)\widehat
I$. Therefore, the average over the exact orbit gives
\begin{equation}
\left\langle\bs{v}(t')\right\rangle\,=\,\exp\left[\frac{1}{2}\int_{t'}^{t}{\rm
    d}\tau_1\int_{t'}^{t}{\rm
    d}\tau_2\left\langle\widehat{\delta\Omega}(\tau_1)
  \widehat{\delta\Omega}(\tau_2)\right\rangle\right]\bs{v}\,=\,
\exp\left[-\frac{2}{3}\tau_{\rm
    c}(t-t')\delta\Omega^2\right]\bs{v}\ .\label{eq:1vc}
\end{equation}
This correlator defines the scattering time $t_{\rm
  s}\,\equiv\,\frac{3}{2}\left(\tau_{\rm
  c}\delta\Omega^2\right)^{-1}$, which depends on the momentum of the
particle. In the following, the generic notation
$C_a(t)\,\equiv\,\exp\left(-a t/t_{\rm s}\right)$ is adopted, with
$a$ a rational number.

The calculation of the correlator $\langle v_i(t_1)v_j(t_2)\rangle$ is
more involved as it involves the product of two time-ordered
exponentials. One must stress that it differs from usual velocity
correlators in diffusion calculations because of the particular
boundary conditions: $v_i(t)=v_i$ and $v_j(t)=v_j$.  This correlator
is written:
\begin{equation}
v_i(t_1)v_j(t_2)\,=\,{\cal T}\exp\left[-\int_{t_1}^{t}{\rm
    d}\tau\,\widehat{\delta\Omega}\right]_i^k \,{\cal
  T}\exp\left[-\int_{t_2}^{t}{\rm
    d}\tau\,\widehat{\delta\Omega}\right]_j^lv_kv_l
\end{equation}
Now, if $t_1>t_2$, one rewrites 
\begin{equation}
{\cal T}\exp\left[-\int_{t_2}^{t}{\rm
    d}\tau\,\widehat{\delta\Omega}\right]_j^l\,=\,
{\cal T}\exp\left[-\int_{t_1}^{t}{\rm
    d}\tau\,\widehat{\delta\Omega}\right]_j^m\,
{\cal T}\exp\left[-\int_{t_2}^{t_1}{\rm
    d}\tau\,\widehat{\delta\Omega}\right]_m^l
\end{equation}
and conversely if $t_2<t_1$. In the following, $\overline
t\,\equiv\,{\rm max}(t_1,t_2)$ and $\underline t\,\equiv\,{\rm
  min}(t_1,t_2)$.  Furthermore, due to the white noise nature of
  $\widehat{\delta\Omega}$,
\begin{eqnarray}
\Biggl\langle{\cal T}&&\exp\left[-\int_{\overline t}^{t}{\rm
    d}\tau\,\widehat{\delta\Omega}\right]\,
{\cal T}\exp\left[-\int_{\overline t}^{t}{\rm
    d}\tau\,\widehat{\delta\Omega}\right]\,
{\cal T}\exp\left[-\int_{\underline t}^{\overline t}{\rm
    d}\tau\,\widehat{\delta\Omega}\right]\Biggr\rangle\,=\nonumber\\
&&\left\langle{\cal T}\exp\left[-\int_{\overline t}^{t}{\rm
    d}\tau\,\widehat{\delta\Omega}\right]\,
{\cal T}\exp\left[-\int_{\overline t}^{t}{\rm
    d}\tau\,\widehat{\delta\Omega}\right]\right\rangle\,
\left\langle{\cal T}\exp\left[-\int_{\underline t}^{\overline t}{\rm
    d}\tau\,\widehat{\delta\Omega}\right]\right\rangle\nonumber\\
&&\label{eq:avprod}
\end{eqnarray}
Finally, the action of $\left\langle{\cal
  T}\exp\left[-\int_{\underline t}^{\overline t}{\rm
    d}\tau\,\widehat{\delta\Omega}\right]\right\rangle$ on $v_kv_l$
gives a factor $C_1\left(\overline t - \underline t\right)v_kv_l$, see
Eq.~(\ref{eq:1vc}). One therefore needs to calculate only the first
average on the r.h.s. of Eq.~(\ref{eq:avprod}).

Here, a key observation is to note that the product of those two time
ordered exponentials can be rewritten as the time ordered exponential
of a tensorial operator; this is demonstrated in
App.~\ref{app:Tordered}.  Expliciting the matrix components:
\begin{equation}
{\cal T}\exp\left[-\int_{\overline t}^{t}{\rm
    d}\tau\,\widehat{\delta\Omega}\right]_i\mbox{}^k \,\,\,{\cal
  T}\exp\left[-\int_{\overline t}^{t}{\rm
    d}\tau\,\widehat{\delta\Omega}\right]_j\mbox{}^l\,=\,{\cal T}
\exp\left[-\int_{\overline t}^{t}{\rm d}\tau\,\delta
    W\right]_{ij}\mbox{}^{kl}\ ,\label{eq:ttW}
\end{equation}
with
\begin{equation}
\delta W_{ij}\mbox{}^{kl}\,=\,\delta W^<\mbox{}_{ij}\mbox{}^{kl}+
\delta W^>\mbox{}_{ij}\mbox{}^{kl},\quad
\delta W^<\mbox{}_{ij}\mbox{}^{kl}\,=\,\delta\Omega_i^k\,\delta_j^l,\quad
\delta  W^<\mbox{}_{ij}\mbox{}^{kl}\,=\,\delta_i^k\,\delta\Omega_j^l
\end{equation}
The tensor product rule in the time ordered exponential is understood
as:
\begin{equation}
\left.\delta  W\cdot\delta
  W\right._{ij}\mbox{}^{kl}\,\equiv\,
\delta  W_{ij}\mbox{}^{mn}\delta
  W_{mn}\mbox{}^{kl}
\end{equation}
 One therefore obtains:
\begin{equation}
\left\langle v_i(t_1)v_j(t_2)\right\rangle\,=\,{\cal T}\exp\left[\frac{1}{2}
\int_{\overline t}^t\int_{\overline t}^t{\rm d}\tau_1{\rm d}\tau_2 
\left\langle \delta W(\tau_1)\cdot\delta
  W(\tau_2)\right\rangle\right]_{ij}\mbox{}^{kl}
C_1(\overline t-\underline t)v_kv_l 
\end{equation}
Now, using the identity
\begin{eqnarray}
\left\langle
\delta\Omega_i\mbox{}^m(\tau_1)\delta\Omega_j\mbox{}^n(\tau_2)\right\rangle&\,=\,&
\left\langle\delta\Omega^a(\tau_1)\delta\Omega^b(\tau_2)\right\rangle\widehat{L_a}_i\mbox{}^m\widehat{L_b}_j\mbox{}^n\nonumber\\
&\,=\,&\frac{2}{3}\tau_{\rm c}\delta(\tau_1-\tau_2)\delta\Omega^2
\left[\delta_{ij}\delta^{mn}-\delta_i\mbox{}^m\delta_j\mbox{}^n\right]
\end{eqnarray}
one finds
\begin{equation}
\left\langle \delta W(\tau_1)\cdot\delta
  W(\tau_2)\right\rangle_{ij}\mbox{}^{kl}\,=\,\frac{2}{3}\tau_{\rm
  c}\delta(\tau_1-\tau_2)\delta\Omega^2\left[-4
  \delta_i\mbox{}^k\delta_j\mbox{}^l -
  2\delta_i\mbox{}^l\delta_j\mbox{}^k +
  2\delta_{ij}\delta^{kl}\right]\ .
\end{equation}
This operator eventually acts on $v_kv_l$, which is symmetric in $k$
and $l$, therefore one can keep only the symmetric part. Define
therefore
\begin{equation}
M_{ij}\mbox{}^{kl}\,=\,\delta_i\mbox{}^k\delta_j\mbox{}^l +
\delta_i\mbox{}^l\delta_j\mbox{}^k - \frac{2}{3}\delta_{ij}\delta^{kl}
\end{equation}
in terms of which one rewrites the symmetrized average, as indicated
by the symbol $(kl)$:
\begin{equation}
\left\langle \delta W(\tau_1)\cdot\delta
W(\tau_2)\right\rangle_{ij}\mbox{}^{(kl)}\,=\,-2\tau_{\rm
  c}\delta(\tau_1-\tau_2)\delta\Omega^2\,M_{ij}\mbox{}^{kl}
\end{equation}

Finally, the $M$ operator satisfies: $M\cdot M\,=\,2 M$ so that
\begin{equation}
\exp(\alpha
M)\,=\,1-\frac{M}{2}+e^{2\alpha}\frac{M}{2}\ ,
\end{equation}
and
\begin{equation}
M_{ij}\mbox{}^{kl}v_kv_l\,=\,2v_iv_j - \frac{2}{3}v^2\delta_{ij}\ .
\end{equation}
Combining together the above results, one ends up with:
\begin{equation}
\left\langle v_i(t_1)v_j(t_2)\right\rangle\,=\,
C_{1}\left(\overline t- \underline t\right)\left\{
C_{3}\left(t-\overline t\right)v_iv_j +\frac{v^2}{3}
\delta_{ij}\left[1-C_3\left(t-\overline t\right)\right]\right\}\ ,
\end{equation}
which has a simple interpretation: the correlator vanishes for time
intervals $\overline t - \underline t$ larger than $t_{\rm s}$, else
it tends towards $v_iv_j$ (the boundary conditions at time $t$) if
$t-\overline t\,\ll\,t_{\rm s}/3$, or to the isotropic average
$\delta_{ij}v^2/3$ in the opposite limit.

One then derives easily the position correlator, with $\Delta r_i(t')
\,\equiv\, r_i(t')-r_i$, including the boundary condition $r_i(t)=r_i$:
\begin{eqnarray}
\left\langle \Delta r_i(t')\Delta r_j(t')\right\rangle\,&=&\, 
\int_{t'}^{t}\int_{t'}^{t} {\rm d}\tau_1{\rm
  d}\tau_2\,\left\langle
v_i(\tau_1)v_j(\tau_2)\right\rangle\nonumber\\
\,&=&\, 2t_{\rm s}^2\Biggl\{\left[\frac{1}{3}
  +\frac{1}{6}C_3(t-t')-\frac{1}{2}C_1(t-t')\right]v_iv_j +
\nonumber\\
&&\,\, \left[\frac{t-t'}{t_{\rm
      s}}-\frac{4}{3}+\frac{3}{2}C_1(t-t')-\frac{1}{6}C_3(t-t')\right]\frac{v^2}{3}\delta_{ij}\Biggr\}\nonumber\\
&&
\end{eqnarray}
Similarly, one obtains
\begin{equation}
\left\langle\Delta r_i(t')\right\rangle\,=\,-t_{\rm
  s}\left[1-C_1(t-t')\right]v_i\ ,
\end{equation}
hence
\begin{eqnarray}
\left\langle\Delta r_i(t')\Delta r_j(t')\right\rangle
-&&\left\langle\Delta r_i(t')\right\rangle\left\langle\Delta
r_j(t')\right\rangle\,=\,\nonumber\\
&&
\quad\quad t_{\rm
  s}^2\Biggl\{\left[-\frac{1}{3}+C_1(t-t')-C_2(t-t')+\frac{1}{3}C_3(t-t')\right]v_iv_j+\nonumber\\
&&\quad\quad\quad\quad +\frac{2}{3}v^2\delta_{ij}\left[\frac{t-t'}{t_{\rm s}}-\frac{4}{3}
  +
  \frac{3}{2}C_1(t-t')-\frac{1}{6}C_3(t-t')\right]\Biggr\}\nonumber\\
&&
\end{eqnarray}
The average $\left\langle\exp\left[i\bs{k}\cdot\bs{\Delta
    r}(t')\right]\right\rangle$ can be truncated at the first
cumulant, leading to
\begin{equation}
\left\langle\exp\left[i\bs{k}\cdot\bs{\Delta
    r}(t')\right]\right\rangle\,\simeq\,\exp\Biggl\{ik^i\cdot\left\langle\Delta
    r_i(t')\right\rangle - \frac{1}{2}k_ik_j\biggl[\left\langle \Delta
  r_i\Delta r_j\right\rangle - \left\langle \Delta r_i\right\rangle
  \left\langle \Delta r_j \right\rangle\biggr]\Biggr\}
\end{equation}
In particular, the small-time limit $t-t'\,\ll\,t_{\rm s}$ will be
useful:
\begin{equation}
\left\langle\exp\left[i\bs{k}\cdot\bs{\Delta
    r}(t')\right]\right\rangle\,\simeq\,\exp\left[-ikv\mu(t-t')
-\frac{1}{6}k^2v^2(1-\mu^2)\left(t-t'\right)^3/t_{\rm
  s}\right]\ ,\label{eq:corrsmallt}
\end{equation}
with $\mu=\bs{k}\bs{v}/(kv)$.

The correlator which enters the expression for the non-linear
susceptibility is 
\begin{equation}
\left\langle\exp\left[i\bs{k}\cdot\bs{\Delta r}(t')\right]
v_j(t')\right\rangle\,=\,\left\langle\exp\left[ i\bs{k}\cdot\bs{\Delta
    r}(t') + \bs{\Delta
    v}\cdot\partial/\partial\bs{v}\right]\right\rangle v_j\ .
\end{equation}
It is understood that $\bs{\Delta v}\,=\,\bs{v}(t')-\bs{v}$ and the
partial derivative $\partial/\partial\bs{v}$ does not act on
$\bs{\Delta v}$. As before, the average of the exponential can be
truncated at the first cumulant, leading to
\begin{eqnarray}
\left\langle\exp\left[i\bs{k}\cdot\bs{\Delta r}(t')\right]
v_j(t')\right\rangle&&\,=\,\nonumber\\ &&
\exp\Biggl\{ik^i\cdot\left\langle\Delta r_i(t')\right\rangle -
\frac{1}{2}k_ik_j\biggl[\left\langle \Delta r_i\Delta r_j\right\rangle
  - \left\langle \Delta r_i\right\rangle \left\langle \Delta r_j
  \right\rangle\biggr]+\nonumber\\ && ik_i\left\langle\Delta
r_i(t')\Delta v_k(t')\right\rangle\partial/\partial v_k-
ik_i\left\langle\Delta r_i(t')\right\rangle\left\langle\Delta
v_k(t')\right\rangle\partial/\partial v_k\Biggr\}v_j\nonumber\\ &&
\end{eqnarray}
where the second-order cumulant $\langle \Delta v_j \Delta
v_k\rangle\partial/\partial v_j\partial/\partial v_k$ has not been
considered because it vanishes when acting on $\bs{v}$. Expanding the
exponential in $\partial/\partial v_j$ to first order, one ends up with
\begin{eqnarray}
\left\langle\exp\left[i\bs{k}\cdot\bs{\Delta r}(t')\right]
v_j(t')\right\rangle
&\,=\,&\left\langle\exp\left[i\bs{k}\cdot\bs{\Delta
    r}(t')\right]\right\rangle \nonumber\\ &&\quad\quad
\times\Biggl\{\langle v_j\rangle + ik_i\biggl[\left\langle \Delta
  r_i(t')\Delta v_j(t')\right\rangle - \left\langle \Delta
  r_i(t')\right\rangle\left\langle \Delta
  v_j(t')\right\rangle\biggr]\Biggr\}\nonumber\\ &\,=\,&
\left\langle\exp\left[i\bs{k}\cdot\bs{\Delta
    r}(t')\right]\right\rangle \Biggl\{
C_1(t-t')v_j+\nonumber\\ &&\quad\quad i kv t_{\rm
  s}\mu\left[\frac{1}{2}C_1(t-t')-C_2(t-t')+\frac{1}{2}C_3(t-t')\right]v_j
-\nonumber\\ &&\quad\quad \frac{i}{3} k_j v^2 t_{\rm
  s}\left[1-\frac{3}{2}C_1(t-t')+\frac{1}{2}C_3(t-t')\right]\Biggr\}
\end{eqnarray}
In calculating the transverse susceptibility, the (longitudinal) last
term $\propto k_j$ disappears, of course.

\section{}\label{app:Tordered}
In order to demonstrate Eq.~(\ref{eq:ttW}), one needs to expand the
time ordered exponentials. The left hand side reads
\begin{eqnarray}
{\cal T}\exp\left[-\int_{\overline t}^{t}{\rm
    d}\tau\,\widehat{\delta\Omega}\right]_i^k &&\,{\cal
  T}\exp\left[-\int_{\overline t}^{t}{\rm
    d}\tau\,\widehat{\delta\Omega}\right]_j^l\,=\,\sum_{m,n=0}^{+\infty}\frac{1}{m!\,n!}\,
\nonumber\\
&& \overbrace{\int_{\overline t}^{t}{\rm d}\tau_1\,
\delta \Omega_{i_1}^{i_2}\int_{\overline t}^{\tau_1}{\rm
  d}\tau_2\,\delta\Omega_{i_2}^{i_3}\ldots}^{m\,{\rm arguments}}
\overbrace{\int_{\overline t}^{t}{\rm d}\tau'_1\,
\delta \Omega_{j_1}^{j_2}\int_{\overline t}^{\tau'_1}{\rm
  d}\tau'_2\,\delta\Omega_{j_2}^{j_3}\ldots}^{n\,{\rm arguments}}\nonumber\\
&&
\end{eqnarray}
and it is understood that $i_1=i$, $i_{m+1}=k$, $j_1=j$,
$j_{n+1}=l$. The product of the $m$ by $n$ integrals can be written as
a single time ordered sequence as follows. For the sake of clarity,
one first rewrites $\delta \Omega^<$ the operators with $i$ indices
and $\delta \Omega^>$ the operators with $j$ indices and one keeps
in mind that all $\delta \Omega^<$ operators are contracted one with
the other according to the time ordered sequence, and similarly for
the $\delta \Omega^>$ operators. Then, one breaks the integral
$\int_{\overline t}^{t}{\rm d}\tau'_1 \delta \Omega^>_{j_1}\mbox{}^{j_2}$ over
the time intervals $[\overline t,\tau_m]$,
$[\tau_m,\tau_{m-1}]$,...$[\tau_1,t]$ and one reorders the sequence,
noting that (indices discarded):
\begin{equation}
\int_{\overline t}^{\tau_i}{\rm d}\tau_{i+1}\,\delta
\Omega^<\int_{\tau_{i+1}}^{\tau_{i}}{\rm d}\tau_{j}\,\delta \Omega^>
\,=\, \int_{\overline t}^{\tau_i}{\rm d}\tau_{j}\,\delta \Omega^>
\int_{\overline t}^{\tau_{j}}{\rm d}\tau_{i+1}\,\delta \Omega^<.
\end{equation}
Repeating this exercise for all $\delta \Omega^>$ integrals, in the
order of the time sequence, one ends up with a time ordered sum over
all possible permutations $\sigma_{mn}$ of the operators:
\begin{eqnarray}
{\cal T}\exp\left[-\int_{\overline t}^{t}{\rm
    d}\tau\,\widehat{\delta\Omega}\right]_i^k &&\,{\cal
  T}\exp\left[-\int_{\overline t}^{t}{\rm
    d}\tau\,\widehat{\delta\Omega}\right]_j^l\,=\,
\sum_{m,n=0}^{+\infty}\frac{1}{m!\,n!}\,\sum_{\sigma_{mn}}\nonumber\\ &&
\int_{\overline t}^{t}{\rm
  d}\tau_1\, \delta \Omega^{\sigma_{mn}(1)}\int_{\overline t}^{\tau_1}{\rm d}\tau_2\,
\delta \Omega^{\sigma_{mn}(2)}\ldots
\int_{\overline t}^{\tau_{m+n}}{\rm
  d}\tau_{m+n}\,\delta\Omega^{\sigma_{mn}(m+n)}\ .\nonumber\\
\end{eqnarray}
The permutation is defined by: $\sigma_{mn}(a)\,=\,\left(>\,{\rm
  or}\,<\right)$, with $m$ copies of $<$ and $n$ copies of $>$. The
indices have been discarded, but it is understood that all operators
are contracted within their respective $<$ or $>$ families, as
mentioned previously. One then notes that this contraction sequence
can be rewritten as the tensor product of $\delta
 W^{<},\,\delta W^{>} $ operators introduced above,
namely:
\begin{equation}
\left[\delta
\Omega^{\sigma_{mn}(1)}\delta\Omega^{\sigma_{mn}(2)}\ldots\delta\Omega^{\sigma_{mn}(m+n)}\right]\mbox{}_{ij}\mbox{}^{kl}
\,=\, \left[\delta W^{\sigma_{mn}(1)}\cdot\delta
W^{\sigma_{mn}(2)}\cdot\ldots\cdot \delta
W^{\sigma_{mn}(m+n)}\right]\mbox{}_{ij}\mbox{}^{kl}
\end{equation}
since $\delta W^<$ acts non-trivially only on $i-$type indices, while
$\delta W^>$ acts non-trivially only on $j-$type indices.

Finally, one uses:
\begin{eqnarray}
\sum_{m,n=0}^{+\infty}\frac{1}{m!\,n!}\,{\cal T}\int{\rm d}\tau_1\,
\delta W^{\sigma_{mn}(1)}&&\int{\rm d}\tau_2\,\delta W^{\sigma_{mn}(2)}\,\ldots
\,=\,
\sum_{p=0}^{\infty}\frac{1}{p!}\sum_{m=0}^{p}\frac{p!}{m!\,(p-m)!}
\nonumber\\
&&{\cal T}\int{\rm d}\tau_1\,\delta W^{\sigma_{m,p-m}(1)}\int{\rm
  d}\tau_2\,\delta W^{\sigma_{m,p-m}(2)}\ldots\nonumber\\
&&
\end{eqnarray}
to obtain
\begin{equation}
{\cal T}\exp\left[-\int_{\overline t}^{t}{\rm
    d}\tau\,\widehat{\delta\Omega}\right]_i^k \,{\cal
  T}\exp\left[-\int_{\overline t}^{t}{\rm
    d}\tau\,\widehat{\delta\Omega}\right]_j^l\,=\,
\sum_{p=0}^{\infty}\frac{1}{p!}\,{\cal T}\left[\int_{\underline
    t}^{\overline t}{\rm d}\tau
  \left(\delta W^<+\delta W^>\right)\right]^p\mbox{}_{ij}\mbox{}^{kl}
\end{equation}
which gives the desired result.

\bibliographystyle{jpp}

\bibliography{shock.bib}

\end{document}